\pdfoutput=1

\documentclass[11pt]{article}

\usepackage[final]{acl}

\usepackage{times}
\usepackage{latexsym}

\usepackage{enumitem}

\usepackage[T1]{fontenc}

\usepackage[utf8]{inputenc}

\usepackage{subcaption}
\usepackage{amsmath}
\usepackage{booktabs}

\usepackage{microtype}

\usepackage{inconsolata}

\usepackage{graphicx}

\title{Improving Citation Text Generation: Overcoming Limitations in Length Control}

\author{Biswadip Mandal, Xiangci Li, Jessica Ouyang \\
  The university of Texas at Dallas, Richardson, TX, USA \\
  \{biswadip.mandal, xiangci.li, jessica.ouyang\}@utdallas.edu
  }

\begin{document}
\maketitle
\begin{abstract}
A key challenge in citation text generation is that the length of generated text often differs from the length of the target, lowering the quality of the generation. While prior works have investigated length-controlled generation, their effectiveness depends on knowing the appropriate generation length. In this work, we present an in-depth study of the limitations of predicting scientific citation text length and explore the use of heuristic estimates of desired length.
\end{abstract}

\section{Introduction}

Citation text generation for scientific articles is the task of summarizing the content of a cited paper as it relates to the citing paper. One key challenge is that the lengths of human-written citations vary greatly. For example, if the citing paper directly builds on the approach of the cited paper, the citation text may be longer and more detailed; if the citing paper mentions the cited paper as one among many similar background works, the citation may be shorter. The challenge for the generation model is to produce a citation that is neither too long, containing irrelevant or redundant information, nor too short, lacking sufficient information.

\begin{figure}[t]
\begin{center}
    \includegraphics[width=0.45\textwidth, height=0.25\textwidth]{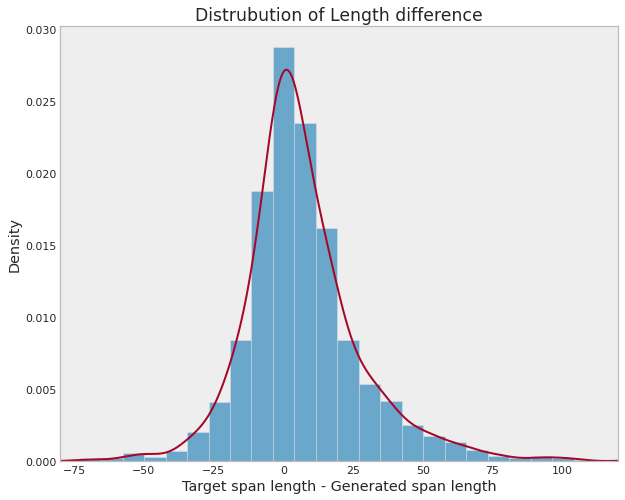}
     \vspace{-0.5em}
  \caption{Length difference in tokens between ground truth citations and \citet{li-etal-2022-corwa}'s generated citations.} 
  \label{fig:generation_length_difference}
  \vspace{-1em}
\end{center}
\end{figure}

In this work, we use the dataset of \citet{li-etal-2022-corwa} and use their annotated \textit{citation spans} as our generation targets. We observe that many of the citations produced by Li et al.'s length-agnostic generation model are of a different length than the human-written targets. Figure \ref{fig:length_difference_example} shows two example citations generated by Li et al.'s model, and Figure \ref{fig:generation_length_difference} shows a histogram of length differences between target citations and Li et al.'s output. 

\begin{figure}[h]
\small
\begin{tabular}{c}
\toprule
    \vspace{.5\baselineskip}
    
    \begin{minipage}[h]{.9\columnwidth}%
    \textbf{\textit{Target}}: Parvez et al. (2018) explicitly model the type of the next word in addition to the word itself. \\
    
    \textbf{\textit{Baseline}}: Parvez et al. (2018) extend vanilla LMs with a language model to model the compatibility between context sentences and labels. However, their approach does not explicitly address the problem of noisy labels.\\

    \textbf{\textit{Oracle}}: Parvez et al. (2018) propose to use LM-enhancement to improve entity typing.\\
    \hrule 
    \vspace{0.5em}
    \textit{\textbf{Target}}: \ldots neural-based methods (Chu et al., 2016\ldots) learn to identify parallel sentences in the semantic spaces. However, these methods require large amounts of parallel sentence pairs\ldots which does not apply to languages with limited resources.\\
    
    \textbf{\textit{Baseline}}: \ldots proposals (Chu et al., 2016\ldots) use the corpora to train models and extract parallel sentences.\\

    \textbf{\textit{Oracle}}: \ldots approaches (Chu et al., 2016\ldots) extract parallel sentences from comparable monolingual corpora using neural networks. However, these methods are only for a small number of languages and do not use any external resources such as dictionaries or corpora. \\
	\end{minipage} \tabularnewline
\bottomrule
    \end{tabular}
    \caption{Examples of generated citations that are too long (top) or too short (bottom).}
    \label{fig:length_difference_example}
\end{figure}

In this work, we experiment with mitigating this length difference by predicting and controlling the length of the generated citation. Controlling generation length in encoder-decoder models has been previously studied in \citet{kikuchi-etal-2016-controlling,fan2017controllable,liu-etal-2018-controlling}. However, predicting the appropriate generation length, rather than using an arbitrary one (eg. a summary length limit), has received less attention. Additionally, what work exists on length prediction \cite{yang-etal-2020-predicting, oka2020incorporating} has been for the task of machine translation, where the lengths of the input and output are correlated because the two must be highly semantically similar. In contrast, for citation generation, the input consists of the abstract of the cited paper and the surrounding context of the target citation, while the output is a concise summary of the cited paper as it relates to the context; the length of the output is not necessarily related to that of the input. 

We tackle the challenge of generating citations that are both informative and of appropriate length. Our main contributions are as follows:
\begin{itemize}[nolistsep,noitemsep]
    \item We demonstrate that controlling the length of generated citation produces a significant improvement in their quality.
    \item We find that automatic citation length prediction is very difficult, and heuristic length estimates give better performance on the downstream citation generation task.
    \item We provide insights into the behavior of generated citations when length is modified.
\end{itemize}  

\section{Related Work}

\textbf{Citation Text Generation.}
Older, extractive approaches select salient sentences from cited papers to serve as citations \cite{hoang2010towards, hu-wan-2014-automatic, chen2019automatic}, while recent work focuses on abstractive approaches to generate citations using neural seq2seq models \cite{xing2020automatic, ge-etal-2021-baco, luu-etal-2021-explaining,chen-etal-2021-capturing,li-etal-2022-corwa}. However, no prior work has investigated the length difference issue addressed in this work, nor attempted length prediction or control for citation text generation.

\textbf{Length Prediction and Control.}
Many works have studied length control \cite{kikuchi2016controlling,fan2018controllable,liu-etal-2018-controlling,lakew2019controlling}, but few have addressed length prediction. For machine translation, where the input and output lengths are related based on their shared semantics, \citet{yang-etal-2020-predicting} trained length prediction and generation jointly, but they used length prediction only as an auxiliary training task to improve translation. \citet{oka2020incorporating} used a length prediction model to control a translation model, but the two were trained separately and used in a pipeline.

\section{Predicting and Controlling Length}

We explore jointly predicting and controlling length for citation generation. We experiment with a multi-task approach where the predicted length controls generation, allowing our length regression model to receive additional training signal from the downstream generation task. We use Longformer Encoder-Decoder \citep[LED;][]{Beltagy2020Longformer} for generation and the length-difference position encoding (LDPE) control approach of \citet{takase2019positional}, which uses the decoder's positional encoding to represent the remaining generation length at each time step (Appendix \ref{subsec:appendix_ldpe}).


\subsection{Automatic Length Prediction}

\begin{figure*}[!ht]
    \centering
    \includegraphics[width=.8\textwidth, height=.25\textwidth]{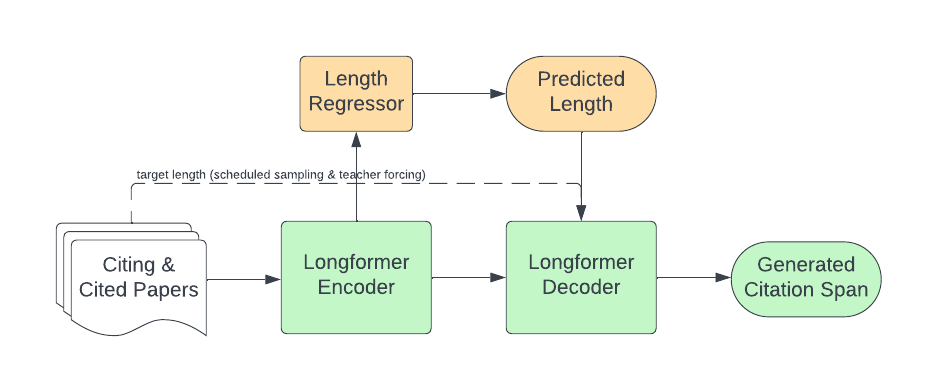}
    \caption{Architecture of our joint length prediction and controlled citation generation models.}
    \label{fig:architecture}
\end{figure*}


Following \citet{li-etal-2022-corwa}, the input $x$ to our system is the concatenation of the citing paper's introduction section; the paragraph containing the (masked) target citation; and the citation mark (eg. ``Smith et al. (2023)"), title, and abstract of the cited paper. To predict the citation length, we take the encoding of the $CLS$ meta-token from the LED encoder and pass it through a feed-forward network ($FF$) to predict a scalar value for length. 
We then use the predicted length to control the length of the generated citation, allowing the length regression model to receive training updates from the downstream generation model (Equation \ref{eq:model}).
\begin{equation}\label{eq:model}
\begin{aligned}
    \mathbf{H} &= LED_{enc}(\mathbf{x}) \\
    \hat{len} &= FF(\mathbf{H}_{<CLS>}) \\
    \hat{y} &= LED_{dec}(\mathbf{H}+\mathbf{LDPE}(\hat{len})) \\
\end{aligned}
\end{equation}

We explore three approaches for training length regression and controlled generation (Figure \ref{fig:architecture}).

\textbf{Vanilla Multitasking.} We jointly train length regression and citation generation using multi-task learning. We pass the predicted length to the generation model for length control and backpropagate the generation loss through the length regression module. The training loss is given by: 
\begin{equation}
    L_{all} = \lambda_{g} L_{gen} +  (1 - \lambda_{g}) L_{len}
\end{equation}
$L_{gen}$ and $L_{len}$ are generation (cross-entropy) loss and regression (root mean square error) loss, respectively. We tried several values for $\lambda_{g}$, but as we discuss in Section \ref{sec:results}, none worked very well; he reported results use $\lambda_{g}=0.3$.

\textbf{Scheduled Sampling Multitasking.} We use scheduled sampling \cite{bengio2015scheduled} to gradually introduce the predicted length to the generation model. We begin by using the ground truth target length with probability $p$, initialized to a high value $p_0$, and slowly decrease it using exponential decay:
\begin{equation}
    p = p_0 * k ^ {(epoch + \frac{step + 1}{ total\_steps})}
\end{equation}
$total\_steps$ is the number of updates per epoch, and $k$ is the decay rate; we use $p_0=0.99$ and $k=0.98$. We do not tune these values, but set them high in order to use the ground truth length for most of training. As we discuss in Section \ref{sec:results}, the vanilla length regression model is not very accurate, so this strategy favors using ground truth length.

\textbf{Teacher Forcing Pipeline} We use a hybrid approach of \citet{yang-etal-2020-predicting,oka2020incorporating}, where length regression is trained as a separate auxiliary task, and its predictions are used only at test time; the generation model is trained using the ground truth length exclusively. 

\subsection{Heuristic Length Estimates}

We also experiment with simple statistical estimates of citation length; we train a separate length-controlled generation model for each heuristic:

\begin{itemize}[nolistsep,noitemsep]
\item \textbf{Average} citation length across the training set.
\item \textbf{Citation marks}: average length across the training set, partitioned by the number of citation marks (i.e. cited papers referenced). 
\item \textbf{Citing paper} average citation length, reflecting the author's style.
\item \textbf{Random} sample from the citation length distribution of the training set.
\end{itemize}

\section{Experimental Settings}
\label{sec:experiments}


We use the CORWA dataset of NLP-domain related work sections annotated with citations; due to the length restriction, we refer the reader to \citet{li-etal-2022-corwa} for details of the dataset.





We compare our experimental approaches to three baselines. First, we train a length control model using the ground truth target citation lengths; this \textit{Oracle} model also uses ground truth lengths at test time, serving as an upper bound on length-controlled citation generation performance. 

Second, the length-agnostic model of \citet{li-etal-2022-corwa} serves as a lower bound that length prediction and control should improve upon. Like our models, Li et al. use LED \cite{Beltagy2020Longformer} with default settings and the same input format.

Finally, we use one-shot prompting with GPT-3.5 Turbo for length-controlled generation; the prompt contains a demonstration example, all inputs used by our models, and the ground truth target citation length. This baseline tests the feasibility of citation length control by prompting LLMs.

\section{Results and Analysis}
\label{sec:results}

\subsection{Length-Controlled Generation}

 \begin{table}[t]
 \small
 \centering
 \begin{tabular}{llcc}
\toprule
 \textbf{Model} & \textbf{R-1} & \textbf{R-2} & \textbf{R-L} \\
\midrule
Oracle (target length) & \textbf{0.295} & \textbf{0.083} & \textbf{0.224} \\ 
\citet{li-etal-2022-corwa} & 0.271 & \textbf{0.079} & 0.212 \\ 
GPT-3.5 Turbo & 0.257 & 0.063 & 0.187 \\ 
\midrule
Vanilla Multitask & 0.236 & \textbf{0.073} & 0.197 \\
Scheduled Sampling & 0.254 & 0.061 & 0.193 \\
Teacher Forcing & \textbf{0.269} & 0.068 & \textbf{0.199} \\
\midrule
H. Average & 0.271 & 0.075 & 0.208  \\
H. Citation Marks & 0.276 & \textbf{0.077} & 0.212   \\
H. Citing Paper & \textbf{0.280} & \textbf{0.077} & \textbf{0.223} \\
H. Random & 0.241 & 0.065 & 0.187 \\
\bottomrule
 \end{tabular}
 \caption{Performance of predicted and heuristic lengths for controlled generation (ROUGE F1).}
 \label{tab:span_generation_results}
 \end{table}

Table \ref{tab:span_generation_results} shows the performance of our multitask approaches to combining length prediction and control, as well as our length estimate heuristics, compared with the baselines.

\textbf{Citation length matters.} The Oracle model achieves the best performance, a significant improvement over Li et al.'s length-agnostic baseline. Further, the Random heuristic has the lowest performance, even worse than baseline. To illustrate the importance of generation length, we see in the first example in Figure \ref{fig:length_difference_example} that the Target does not criticize the cited paper, and the Baseline erroneously does so; the Oracle citation fixes this problem by eliminating the extra sentence. An opposite case can be seen in the second example.  

\textbf{Citation length is idiosyncratic.} Of our approaches, the Citing Paper heuristic performs best, suggesting that citation lengths within a paper do not vary as much as citations from different papers. This may explain why length prediction is so challenging; citation length depends on the author. 


\subsection{The Effects of Length Prediction}

To examine why joint length prediction and control was outperformed by the Citing Paper heuristic, we use mean absolute error (MAE) to measure our length regression performance.
We evaluate the ability of the generation model to control output length using \textit{Control Variance} \cite{liu-etal-2018-controlling}:
\begin{equation}\label{eq:length_var_measure}
    Control\_Var = 0.001 * \frac{1}{n} \sum_{i=0}^{n} |l_i - len_i| ^ 2
\end{equation} 
$n$ is the number of datapoints and $l_i$ and \textit{$len_i$} are the generated and desired length, respectively.

\begin{table}[t]
\centering
\small
\begin{tabular}{lc c}
\toprule
\textbf{Model} & MAE & Control Variance \\
\midrule
Oracle & - & \textbf{0.0001} \\
GPT-3.5 Turbo & - & 0.1852 \\
\midrule
Vanilla Multitask & 15.04 & 0.5170 \\
Scheduled Sampling & \textbf{13.45} & 0.0138 \\
Teacher Forcing & 13.84 & \textbf{0.0051} \\ 
\bottomrule
\end{tabular}
\caption{Performance on length regression and control. Length regression is evaluated using MAE in number of tokens; control is evaluated using variance between desired and generated length.}
\label{tab:length_prediction}
\end{table}


\textbf{Citation length prediction is hard.} We find that all of our length regression models struggle to accurately predict citation length. Table \ref{tab:length_prediction} shows that all three approaches have MAE of 13 tokens or more, which is significant, considering the average length of citations in the dataset is only 34.5 tokens. We do find that the Vanilla and Scheduled Sampling strategies, which are updated with the generation loss, perform better than Teacher Forcing, but overall their performance is poor and causes the downstream generation task to perform worse than Baseline. Further, we see that, while the Oracle approach exhibits a high level of controllability, our multitask strategies do not. This gap may arise from our predicted length being so noisy: the downstream length control model learns to ignore this unreliable length signal.

To further investigate the impact of input length errors, we conduct an experiment where we take the same target citation and generate it at different lengths: 20, 30, or 50 tokens. We find that forcing the model to generate short citations results in vague, generic citations. 
Increasing the length results in additional sentences being generated, which can comment about the methodology, add more information, or criticize the cited paper. However, when the desired length exceeds the ground truth, the model begins to hallucinate these comments, especially criticisms. Figure \ref{fig:length_effect2} shows a citation being generated at both the correct (20 tokens) and a longer (50 token) length. The longer generation hallucinates a criticism that clearly contradicts the cited paper's abstract. 

Finally, we see that GPT-3.5 has relatively poor length controllability, corroborating the observations of \citet{sun2023evaluating} regarding the management of output length in LLMs through prompting.

\begin{figure}[t]
\small
\begin{tabular}{l}
\toprule
\vspace{.5\baselineskip}
\begin{minipage}[h]{.9\columnwidth}%

\textbf{len=20}: \ldots Wang and Lu (2018a) proposed a segmental hypergraph model for modeling. \\ \\
\textbf{len=50}: \ldots Wang and Lu (2018a) proposed a segmental hypergraph representation to model overlapping entities. However, their model is \textit{not able to capture the interactions} between entities with overlapping spans. \\ \\
\textbf{Wang and Lu (2018a) abstract}: \ldots We show that our model built on top of such a new representation is \textit{able to capture features and interactions} that cannot be captured by previous models\ldots \\
\end{minipage}\tabularnewline
\bottomrule
\end{tabular}
\caption{Example of over-long citation resulting in a \textit{hallucinated} criticism of the cited paper.}
\label{fig:length_effect2}
\end{figure}

\section{Conclusion}

We have explored different approaches for predicting the lengths of citations and using length to control generation. Our experimental results demonstrate that using the ground truth length can significantly improve the quality of the generated citations, but predicting the length of a citation based solely on the cited paper abstract and target citation context is challenging. To address this issue, we propose the use of heuristic estimates of desired length, finding that citation length is mostly related to each author's individual writing style, which explains why it is so difficult to predict. Our work highlights an important challenge in citation text generation and suggests a straightforward solution. By using author-specific citation length estimates, we can eliminate the length difference between the generated and ground truth citations, significantly improving the quality of the generated citations. We have made our code available at \url{https://github.com/mandalbiswadip/LengthControlledGeneration}




\bibliography{acl_latex, anthology}
\bibliographystyle{acl_natbib}

\appendix

\section{Appendix}
\label{sec:appendix}

\subsection{Length-Difference Positional Encoding (LDPE)}
\label{subsec:appendix_ldpe}

\citet{takase2019positional} define the LDPE for a Transformer-based encoder-decoder as follows: 
\begin{equation}\label{eq:ldpe}
\begin{aligned}
    LDPE (pos, len, 2i) &= \sin{ ( \frac{(len - pos)}{10000^ \frac{2i}{d}} )} \\
    LDPE (pos, len, 2i + 1) &= \cos{ ( \frac{(len - pos)}{10000^ \frac{2i}{d}} )} \\
\end{aligned}
\end{equation}
$len$ refers to the desired length of the generation, measured in the number of tokens.

\subsection{More Citation Length Examples}
\label{subsec:length_effect}

Figure \ref{fig:length_effect1} shows an example of a citation that is vague and too generic when generated at a shorter length. Figure \ref{fig:length_effect3} shows two examples of citations generated at shorter or longer lengths, resulting in lower fluency. 

\begin{figure}[ht]
\small
\begin{tabular}{l}
\toprule
\vspace{.5\baselineskip}
\begin{minipage}[h]{.9\columnwidth}%
\textbf{len=20:} Peng et al. (2018) improve the state-of-the-art. \\

\textbf{len=30:} Peng et al. (2018) propose a joint model for frame-semantic parsing and semantic dependency parsing by treating annotations as variables.\\

\end{minipage}\tabularnewline
\bottomrule
\end{tabular}
\caption{Example of an overly-short, generic citation.}
\label{fig:length_effect1}
\end{figure}

\begin{figure}[ht]
\small
\begin{tabular}{l}
\toprule
\vspace{.5\baselineskip}
\begin{minipage}[h]{.9\columnwidth}%
\textbf{\textit{Too short:}} Alkhouli et al. (2018) proposed to use an additional alignment head to improve NMT performance, but their scalability. \\

\textbf{\textit{Too long:}} Firat et al. (2016b) proposed a distillation-based distillation approach, in which the distillation process is trained on a small set of labeled data. \\
\end{minipage}\tabularnewline
\bottomrule
\end{tabular}
\caption{Examples of shorter and longer generated citations (compared to the ground truth target length) resulting in poor fluency.}
\label{fig:length_effect3}
\end{figure}

\subsection{GPT-3.5 Turbo Prompt}
\label{subsec:gpt_prompt}

Figure \ref{fig:gpt_prompt} shows the prompt we used for our GPT-3.5 Turbo length control baseline. The prompt consists of the citation context and a list of papers to cite, as well as the ground truth target citation length.

\begin{figure}[t]
\small
\begin{tabular}{l}
\toprule
\vspace{.5\baselineskip}
\begin{minipage}[h]{.9\columnwidth}%
    We have written an incomplete related work section. We want a set of cited papers to be summarized in the context of the imcomplete related work section. Given the incomplete related work section and a list of cited papers, complete it by summarizing the cited papers in the related work section context in <Length> words. The position where the summarized text goes is indicated by the special token `[Dominant]'. \\
\hrule
\vspace{\baselineskip}

    The cited paper can be one of two types, referred to as Citation Type: 
    
    Dominant: These citations are discussed in detail, usually via summarization of their content, and are often longer than reference citations.
    
    Reference: These citations are not discussed in detail. Reference citations tend to be more abstract than dominant citations. \\
\hrule
\vspace{\baselineskip}

    Output format: Summarize the cited papers given their Citation Mark, Citation Type, Title, and Abstract. Return only the piece of text required to complete the related work section. Return it in string format. The number of words in the output should be the same as <Length>. \\
\hrule
\vspace{\baselineskip}

    Incomplete Related work section: 
    <Example 1 incomplete related work section>

    List of cited papers:

    1.

    Citation mark: <Example 1 first cited paper citation mark>

    Citation Type: <Example 1 first cited paper citation type>

    Title: <Example 1 first cited paper title>

    Abstract: <Example 1 first cited paper abstract>

    2.

    \ldots

    Number of output words: <Example 1 Length>

    Output:
    ```<Example 1 output>'''

    Explanation: The generated output has <Example 1 Length> words, which starts with the word <Example 1 output starting word> and ends with the word <Example 1 output ending word> \\
\hrule
\vspace{\baselineskip}

    Incomplete Related work section:
    <Input incomplete related work section>

    List of cited papers:

    1.

    Citation mark: <Input first cited paper citation mark>

    Citation Type: <Input first cited paper citation type>

    Title: <Input first cited paper title>

    Abstract: <Input first cited paper abstract>
    
    2.

    \ldots

    Number of output words: <Length>

    Output:
\end{minipage}\tabularnewline
\bottomrule
\end{tabular}
\caption{One-shot prompt used for length-controlled citation generation using GPT-3.5 Turbo.}
\label{fig:gpt_prompt}
\end{figure}

\end{document}